# Understanding the degradation of a model Si-anode in Li-ion battery at the atomic-scale

Se-Ho Kim[1,†,*], Kang Dong[2,†], Huan Zhao[1], Ayman A. El-Zoka[1], Xuyang Zhou[1], Eric V. Woods[1], Finn Giuliani[3], Ingo Manke[2], Dierk Raabe[1], Baptiste Gault[1,3,*]

[1] Max-Planck Institut für Eisenforschung GmbH. Max-Planck-Straße 1, 40237 Düsseldorf, Germany
[2] Institute of Applied Materials, Helmholtz-Zentrum Berlin für Materialien und Energie. Berlin, 14109 Germany
[3] Department of Materials, Royal School of Mines, Imperial College. London, SW7 2AZ, United Kingdom

[†]these authors contributed equally

*Corr. Authors: s.kim@mpie.de, b.gault@mpie.de

**Abstract**

Si-anodes have long been candidates thanks to an expected ten-fold increase in capacity compared to graphite. However, details of the mechanisms governing their degradation remain elusive, hindering science-guided development of long-lived Si-based anodes. Here, to advance the understanding of the degradation of the electrolyte and electrode, and their interface, we exploit the latest developments in cryo-atom probe tomography to study a model, single crystal Si anode during cycling. We evidence anode corrosion from the decomposition of the Li-salt before charge-discharge cycles even begin. The newly created grain boundaries facilitate pulverization of nanoscale Si fragments, one is found floating in the electrolyte. As structural defects are bound to assist the nucleation of Li-rich phases in subsequent lithiations and accelerate the electrolyte's decomposition, these insights into the developed nanoscale microstructure interacting with the electrolyte contribute to understanding the self-catalysed/accelerated degradation Si-anodes and can inform new battery designs unaffected by these life-limiting factors.

**Keywords:** atom probe tomography • silicon anode • lithium ion battery • degradation

To meet the rapidly increasing demand for Li-ion batteries for electric vehicles[1,2], tremendous efforts have been devoted to discover cheap and abundant anode material that can replace graphite that is in short supply[3]. A crystalline Si anode, which can offer nearly ten times the capacity of a commercial graphite anode ($Q_{Si}$ = 3600 mAh $g^{-1}$ *vs.* $Q_{graphite}$ = 372 mAh $g^{-1}$)[4], has emerged as an attractive anode material for next-generation Li-ion batteries since the first development of the Li-Si anode by Lai in 1976[5]. Compared to graphite, in which each of the six in-plane C atoms can only bond with one Li ion, each Si atom can bond with up to 4.4 Li ions[6]. Thus, finding a path to exploiting Si as anode material can be a revolutionary approach for reaching batteries with ultra-high energy density. Telsa Inc. has revealed its plans to gradually increase the use of Si anode in its future batteries[7], and Amprius Tech. Inc. recently announced the shipment of its first commercially available Li-Si battery cells with energy density of 450 mWh $g^{-1}$ [8].

An efficient Si anode remains some sort of *holy grail* for rechargeable Li-ion batteries, and their widespread use is hindered by rapid capacity fading[9,10]. The enormous volume changes occurring during lithiation/delithiation cycles (*e.g.* +300 % volume increase from Si to $Li_{22}Si_5$) result in irreversible damage[4]: deformation and residual stresses accumulate and create an ensemble of structural defect features and their respective chemical decoration states, including interfaces, dislocations, grain boundaries, and nano-cracks forming within the Si anode. An array of approaches has been explored to overcome this critical issue. For example, the use of nano-composite/structured Si[11], including nanowires[12,13], core-shell[14,15] and hollow[16,17] nanoparticles, and porous Si[18,19], has been reported to be effective for the enhanced suppression of the initiation of mechanical fracture from the large volume changes.

Most studies use techniques providing a bulk average or two-dimensional information[20–22], which, even in combination, cannot analyse the nanoscale compositional distribution and microstructural evolution of electrodes and electrolytes. In-situ[11,23,24] and cryogenic transmission electron microscopy (TEM)[25,26] have already revealed an undesirable removal or destruction of the passivating solid-electrolyte interphase (SEI), and severe pulverization of Si nano/micro-particles from bulk Si during the expansion and shrinkage cycles[26]. Despite impressive empirical advances, numerous fundamental aspects of the microstructural degradation hence remain elusive, making it impossible to devise targeted strategies to circumvent these specific issues and enable a breakthrough in Si-based anodes. Elucidating the deformation that led to mechanical failure has emerged as a crucial topic to achieve a high-capacity Si anode.

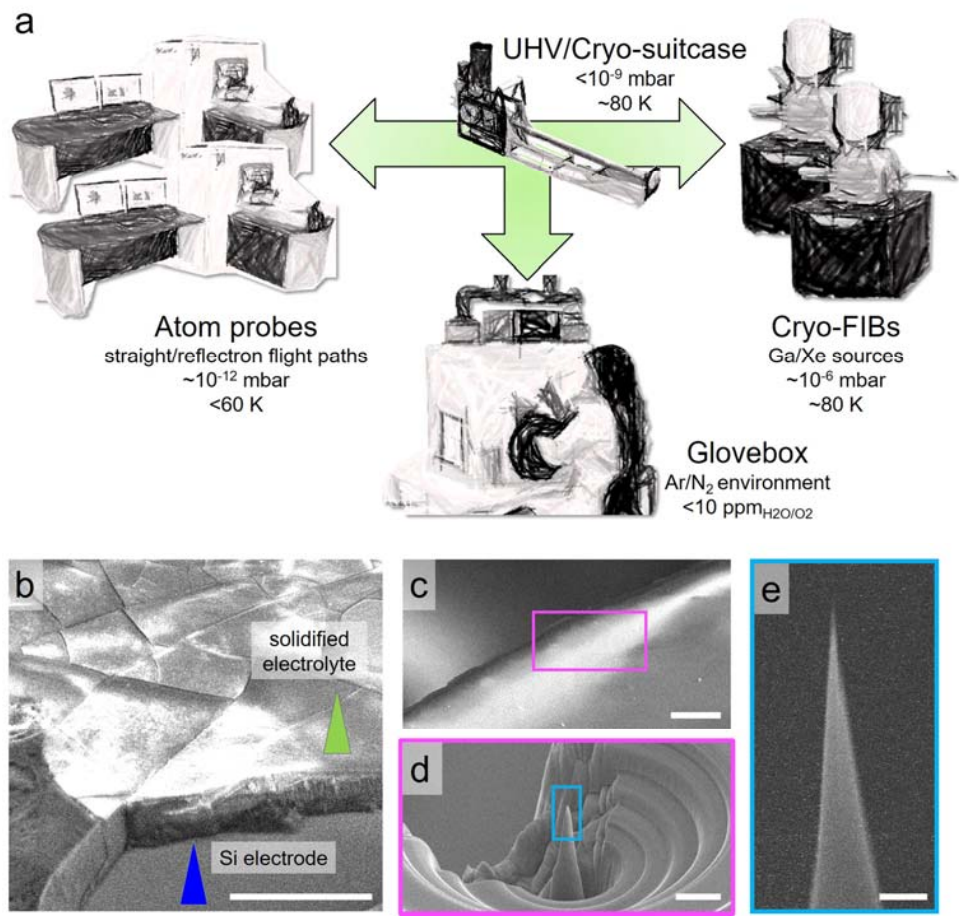

**Figure 1.** a) Unique infrastructure for cryo-atom probe enabling the study. (b) SEM images of the LN2-quenched anode containing the frozen-electrolyte surface; (c) the Si electrode where the (d) cryo-milled pillar was made to prepare the (e) final APT specimen. Scale bars are 50µm in (b)&(c), 20 µm in (d) and 1 µm in (e).

The combination of high-resolution microscopy of the fine scale of the microstructure that develops, and the precise microanalysis of the electrode's evolving composition, can be achieved by using the latest development in cryogenic atom probe tomography (cryo-APT), Figure 1a. APT provides direct and three-dimensional, near-atomically resolved analytical imaging of materials and has the ability to collect all element irrespective of their mass. APT is underpinned by an intense electric field that provides controlled removal of individual ions from a sharp, needle-shaped specimen. However, this field can cause outwards electromigration of Li[27] in battery materials, making impossible the detailed analysis of its distribution but also affecting the overall data quality[28], which explains why battery materials have rarely been analysed by APT[29–32]. However, we demonstrated recently approaches enabling analysis of lithiated anode and cathode materials[28], and delithiated samples still bear traces of crucial processes taking place during battery operation.

Here, we leverage cryo-APT for the first time to obtain compositional mapping of Li-ion battery materials, the abutting electrolyte and the solid-liquid interface between the two at increasing number of charge-discharge cycles. Custom cells were disassembled inside a $N_2$ glovebox ($H_2O$ and $O_2 < 10ppm$)[33], see Methods. The collected Si anode with the electrolyte were immediately plunge-frozen in liquid-$N_2$ (LN$_2$), then transferred by using the cryogenically-cooled, ultra-high-vacuum suitcases into a scanning-electron microscope / Xe-plasma focused ion beam (SEM/PFIB) for imaging and cryogenic specimen preparation, Figure 1b–c. APT specimens of the electrolyte and electrode were prepared at cryogenic temperature using the method we introduced in Ref.[34], Figure 1d–e.

The location of the cryo-APT analyses of the uncycled electrode and electrolyte, Figure 2c, are indicatively marked in Figure 1b. Within the electrolyte, individual, isolated Si ions are already detected. We conducted a cryo-APT analysis of the frozen raw electrolyte on a different metallic substrate (Au) that shows no Si ion (see Figure S1-S4). These additional analyses confirm that dissolved Si ions originated from the corrosion of the Si anode. Veith *et al.* observed non-electrochemically driven Si-O and Si-F bonds on a Si anode soaked in a similar electrolyte[35]. Si-O groups can react with HF generated by hydrolyzed or thermally-decomposed LiPF$_6$ electrolyte[36], resulting in the dissolution of Si ions and two additional $H_2O$ molecules, which trigger further HF generation and a self-sustaining corrosive cycle[37–39]. The hydrolysis could be initiated by residual atmospheric moisture during cell assembly[40]. Degradation of the anode and the electrolyte hence start even before cycling, with any oxygen-containing Si species that generate more water and accelerate the failure of the Si battery cell.

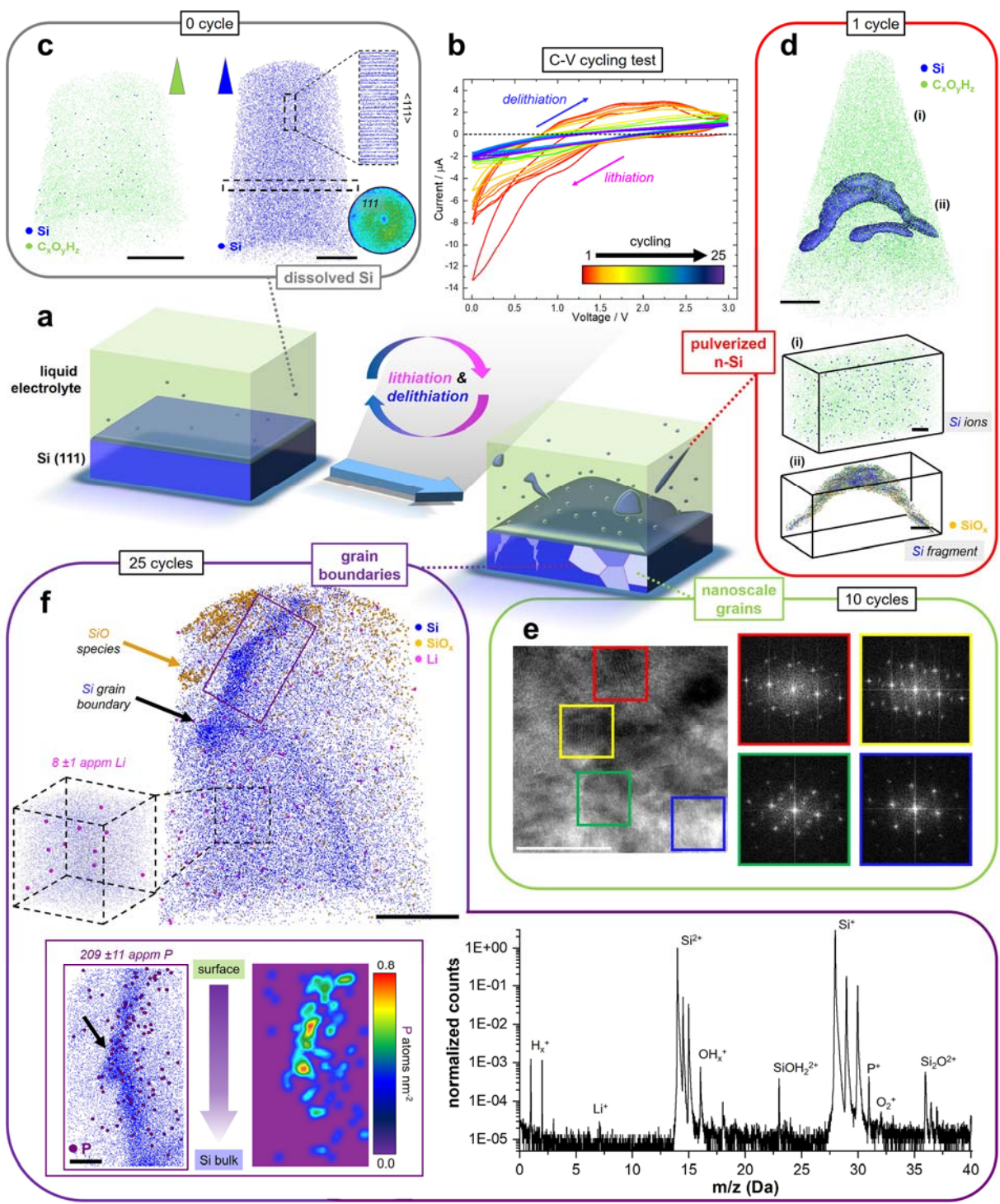

**Figure 2.** (a) Schematic of the Si electrode and cycling process. (b) Voltage vs. current curves of the Si(111) anode in a Li-Si cell. (c) cryo-APT analysis of the electrolyte and anode before cycling; scale bars are 20 nm. (d) cryo-APT reconstructed atom map of the 1-cycle electrolyte; the blue iso-surface delineates regions containing at least 25 at.% Si; scale bar is 20 nm. Movie (#1), the corresponding mass spectra and additional analyses can be found in SI; (i) is a close-up showing dissolved Si ions (scale bar = 2 nm) and (ii) is a delaminated Si debris in the electrolyte (scale bar = 5 nm). Green, blue, and yellow dots represent reconstructed carbonate species, Si and SiOx compounds, respectively. (e) transmission-electron

micrograph of the 10-cycled Si anode along the [110] zone axis of the single-crystal, along with Fast Fourier Transformation (FFT) patterns from different regions highlighted by coloured boxes. A white scale bar is 20 nm. (f) 3D reconstructed atom map of the Si electrode after 25-cycles (scale bar = 20 nm). Blue, yellow, and pink dots represent reconstructed Si, SiOx, and Li, respectively. Movie (#2) and mass spectra of corresponding dataset are presented in the SI. Inset shows the extracted Si grain boundary with the 2D contour density map of P atoms (scale bar = 5 nm).

After 1 cycle, Figure 1d, cryo-APT reveals also dissolved isolated Si ions, accompanied by an approx.10-nm pulverized Si fragment covered with an oxide shell (see Figure S5 and S6). Such a fragment could potentially block the pores of the separator for Li-ion diffusion, raising cell impedance and deteriorating the rate performance of the battery. The presence of the $SiO_x$-species at the surface support the hypothesis that the dissolution was associated to the formation of HF.

Already after 10-cycles, TEM was performed on the dried electrode after removal of the electrolyte and thorough cleaning, Figure 2e, complemented by additional APT experiments (Figure S7-S13). We evidences that the originally single crystalline Si has transformed into a nano-crystalline microstructure, containing numerous nanoscale grains and grain boundaries with different crystallographic orientations that have been formed during the lithiation/delithiation process, confirming previous reports[41]. The volume change associated with the formation of Li-rich phases imposes strong compressive loading on the silicon matrix[42]. Indentation of Si single-crystals has demonstrated that the breaking of covalent Si-Si bonds injects a high number of vacancies in the crystal and results in amorphization[43,44], also recently reported experimentally during battery cycling[45]. Upon relaxation during delithiation, depending on the rate, new crystals nucleate with no orientation relationship with the surrounding crystal matrix[43,44]. The discharging rate must influence this process.

After 25 cycles, Figure 2f, cryo-APT analysis of the very surface of the anode contains two Si grains, as confirmed by atom probe crystallography[46] (Figure S14), and a faceted grain boundary. No chemicals expected from the SEI layer ($LiCO_2$, LiOH, LiF) are observed at the interface (see Movie #3), which can be attributed to the high reversibility of the SEI layer on Si anodes[25,47,48]. On the surface we find several isolated nanoscale islands rich in $SiO_x$ species. Such oxides promote the formation of HF which corrodes the Si, passivate the Si anode and act as a mechanical clamping layer that restricts swelling[48,49]. $SiO_x$ can store Li ions ($Q_{SiO}$ = 1543 mAh g$^{-1}$)[50] with lower volume expansion (approx. 120%) when irreversibly lithiated[51], that can cause stress build-up at the interface and facilitate crack initiation[52], decohesion and pulverization, explaining the presence of SiO$_x$ on the Si fragment's surface in Figure 2d.

After full delithiation, 20-30nm below the surface, Li (8 ±1 appm) is still detected within the Si matrix, as readily visible from the corresponding mass spectrum, Figure 2f. Density-functional theory predicts an attraction between vacancies and Li in Si[53], which can combine with a strong Coulomb attraction between an electron-rich vacancy and the electropositive Li. The image Li atoms are hence likely trapped by remaining vacancies injected under plastic loading.

At the grain boundary, Li does not appear segregated, conversely to P, that is even seen partitions to specific facets and to the facet junction indicated by a black arrow (see Figure S15). This distribution was previously suggested to be associated to local strain[54]. P can diffuse along grain boundaries in Si[55], and its segregation can be energetically favourable due to the passivation of dangling bonds[56], which modifies the conductivity[57]. In addition, atomistic simulations have indicated that the combined effect of the presence of P and a stress concentrator (i.e. a grain boundary) decreases the fracture strength of Si-nanowires[58]. Lastly, the presence of P (209 ±11 appm), originally from the LiPF$_6$ salt, also suggests the liberation of F and the facile formation of HF that is a known embrittler of polycrystalline-Si through void formation along grain boundaries[59]. These effects collectively make these newly created grain boundaries particularly brittle and critical to the lifetime of the Si-anode.

To summarize, cryo-APT allowed us to track the evolution of the three-dimensional, nanoscale elemental distributions of species in the electrolyte, a model Si anode and their interface over increasing charge-discharge cycles. We provide measured data that advance the understanding of the degradation mechanism – or actually degradation mechanisms – and emphasise the often-overlooked role of microstructural defects created and evolving throughout the battery operation lifetime. In addition, the nucleation of the Li$_x$Si$_y$ (metastable) phases[42] during the first cycle can be assumed to be homogeneous, occurring randomly across the surface of the anode. However, the combined presence of crystalline defects and remaining Li-impurities in the anode will undoubtedly assist heterogeneous nucleation of these phases during subsequent lithiation, potentially enhanced by accelerated diffusion of Li through structural defects[60,61]. Segregants can energetically destabilize grain boundaries, already weakened by HF[59], or form space charges, that can favour decohesion. Nucleation in the parts of the microstructure with a high density of defects localizes the volume expansion to mechanically weaker regions, thus facilitating the pulverization of fragments from the anode. This combination of (electro)chemical reactions, phase transformation, and mechanical failure, assisted by the localised decomposition of the electrolyte, accelerates the delamination/mass loss and localized lithiation causing fast loss of capacity[51] (see Figure S16). Strategies for the development of robust and durable Si-based anodes for next-generation Li-ion batteries can draw from our findings on the degradation of Si electrode –the role of the newly formed grain boundaries that may be exploited through segregation, but also the details of the electrolyte degradation that can guide the selection of P-free and F-free salts and avoiding exposure to moisture during fabrication, which can be difficult to achieve in large-scale production.

**Acknowledgements**

BG is grateful for fruitful discussions and insights from Dr. Christoph Freysoldt. SHK, EW, AEZ, BG are grateful for financial support from the ERC-CoG-SHINE-771602 at some point in the past few years. SHK, AEZ, DR, BG acknowledge financial support from the DFG through the DIP Project number 450800666. XZ is grateful for financial support from the Alexander von Humboldt Foundation.

# Supplementary Information

# Understanding the degradation of a model Si-anode in Li-ion battery at the atomic-scale

**Authors:** Se-Ho Kim[1,†,*], Kang Dong[2,†], Huan Zhao[1], Ayman A. El-Zoka[1], Xuyang Zhou[1], Eric V. Woods[1], Finn Giuliani[3], Ingo Manke[2], Dierk Raabe[1], Baptiste Gault[1,3,*]

**Affiliations:**

[1]Max-Planck-Institut für Eisenforschung GmbH, Max-Planck-Straße 1, 40237 Düsseldorf, Germany

[2]Institute of Applied Materials, Helmholtz-Zentrum Berlin für Materialien and Energie, 14109 Berlin, Germany

[3]Department of Materials, Royal School of Mines, Imperial College, SW7 2AZ London, United Kingdom

†co-first authors

*corresponding authors

# Methods
## Materials

Silicon wafer (0.5 mm thick, no dopant, (111)) and an electrolyte of 1 M LiPF$_6$ with a mixture of ethylene carbonate (EC) and diethyl carbonate (DEC) (1:1, v/v) were received from Sigma-Aldrich. Metallic lithium was purchased from MTI Corp. USA. A Swagelok derived cell consisting of a polyether ether ketone (PEEK) housing is described in our previous report[67].

## Battery Assembly and Cycling

Si wafer was cut into 3.0 mm disks by laser-cutting under Argon gas protection. The obtained Si disks were washed using ultrapure water (Milli-Q) and 2-propanol (HPLC Plus, 99.9%, Sigma-Aldrich) to remove contaminations (*e.g.* dust) on the surface from the cutting process. Li chips were punched into disks with a diameter of 3 mm as the counter electrode without further treatment. To preserve the SEI layer on the surface of Si wafer, instead of using a typical Celgard or glassfiber separator, we opt for a ring-shaped spacer (0.5 mm thick) made of polytetrafluoroethylene (PTFE) as a separator. In this way, the Si surface including the SEI layer at the middle area of the Si disk could keep intact without undergoing a peeling-off procedure during the cell disassembly. The Li/Si cells were built using the customized Swagelok-type cell in an argon-filled MBraun glovebox (H$_2$O and O$_2$ < 5 ppm). After cell assembly, Li/Si cells were subjected to cyclic voltammetry (CV) cycling between 0.01 – 3.0 V at a scan rate of 0.2 mV/s using a BioLogic MPG-200 potentiostat. All the cells were stopped after 0, 1, 10, 25 cycles at the delithiation state (see Fig. S17).

## Sample preparation

After cycling, the cell was disassembled in a nitrogen filled glove-box (<10 ppm H$_2$O and O$_2$). Without washing, the Si disk was mounted on a Cu clip and was rapidly plunged into liquid nitrogen, followed by loading on a scanning-electron microscope/Xe-plasma focused ion beam (SEM/p-FIB) (Helios PFIB, Thermo-Fisher, Eindhoven, Netherlands) stage. Subsequently, the Si disk was transferred into the cryo-p-FIB chamber using the ultra-high vacuum transfer suitcase ($10^{-9}$ mbar, -190 $^{o}$C) (VSN-40, Ferrovac GmbH, Zurich, Switzerland) to avoid the sample exposure to air. An illustration of environmentally sensitive sample preparation/transfer for FIB/APT is shown in Fig. S18.

## Cryo-APT specimen preparation

The cryo-p-FIB stage (Gatan C1001, Gatan Inc., California, USA) was pre-cooled to -190 $^{o}$C by cold $N_2$ gas. A clean pillar from frozen electrolyte and Si anode was prepared using the in-situ non lift-out protocol described in references[32,68]. After the height of the post had reached 50 μm, progressively the frozen sample were sharpened into APT specimen using annular milling patterns (e.g. specimen-radius less than 100 nm). Scanning electron micrographs were taken at 5-15 kV and 1.6-2.3 nA to avoid charging effects and the e-beam-induced diffusion/reaction (see Fig. S19 and S20).

## APT measurement

Atom probe data were acquired from 5000 XS instrument (CAMECA, Madison, USA) in pulsed laser mode with laser energy of 20-40 pJ for the cold Si anode (80 pJ for the frozen electrolytes) and rate of 100 kHz at 1 % detection rate. The base temperature was set to 60 K throughout the measurement and the applied direct current was adjusted to control the stable evaporation (see Fig. S21). The atom map reconstruction and data analysis were done using AP SUITE 6.1 software developed by CAMECA.

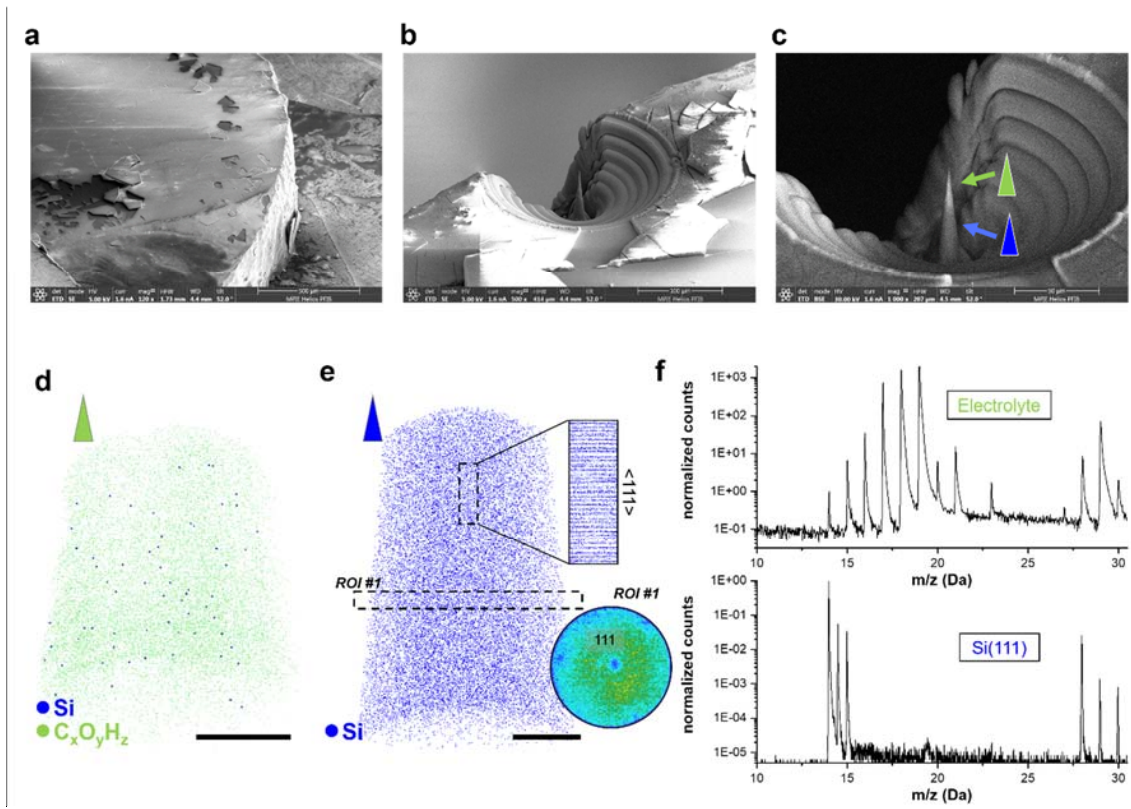

**Fig. S1.** (a) SEM image of frozen 0-cycle Si anode. (b) In-situ annular milling process. (c) Final APT specimens. 3D atom maps of the 0-cycle specimen: (d) electrolyte and (e) single crystal Si(111) anode. Scale bars are 20 nm. (f) corresponding mass spectrum of each. Here in the electrolyte mass-spectrum, no Li peaks were measured; however, we detected a strong peak at 19-21 Da which could be originated from $LiC^+$ (or $F^+$) and $LiCH_x^+$ (or $H_xF/LiO_x^+$) species. Nevertheless, herein, no segregation behavior of carbonate species nor no $LiPF_6$ salt are detected. The decomposed C:O atomic ratio from the acquired mass spectrum is 1.07 which supports that the frozen specimen is the electrolyte compound. The molecular formulae of mixed organic solvent of ethyl carbonate (EC) and dimethyl carbonate (DMC) compounds are $C_3O_3H_4$ and $C_3O_3H_6$, respectively, so it is difficult to conclude which molecules they are. Apart from that, as expected, the atom map of the as-received single-crystal Si anode shows [111] crystallographic pole at the center with Si(111) atomic planes readily visible.

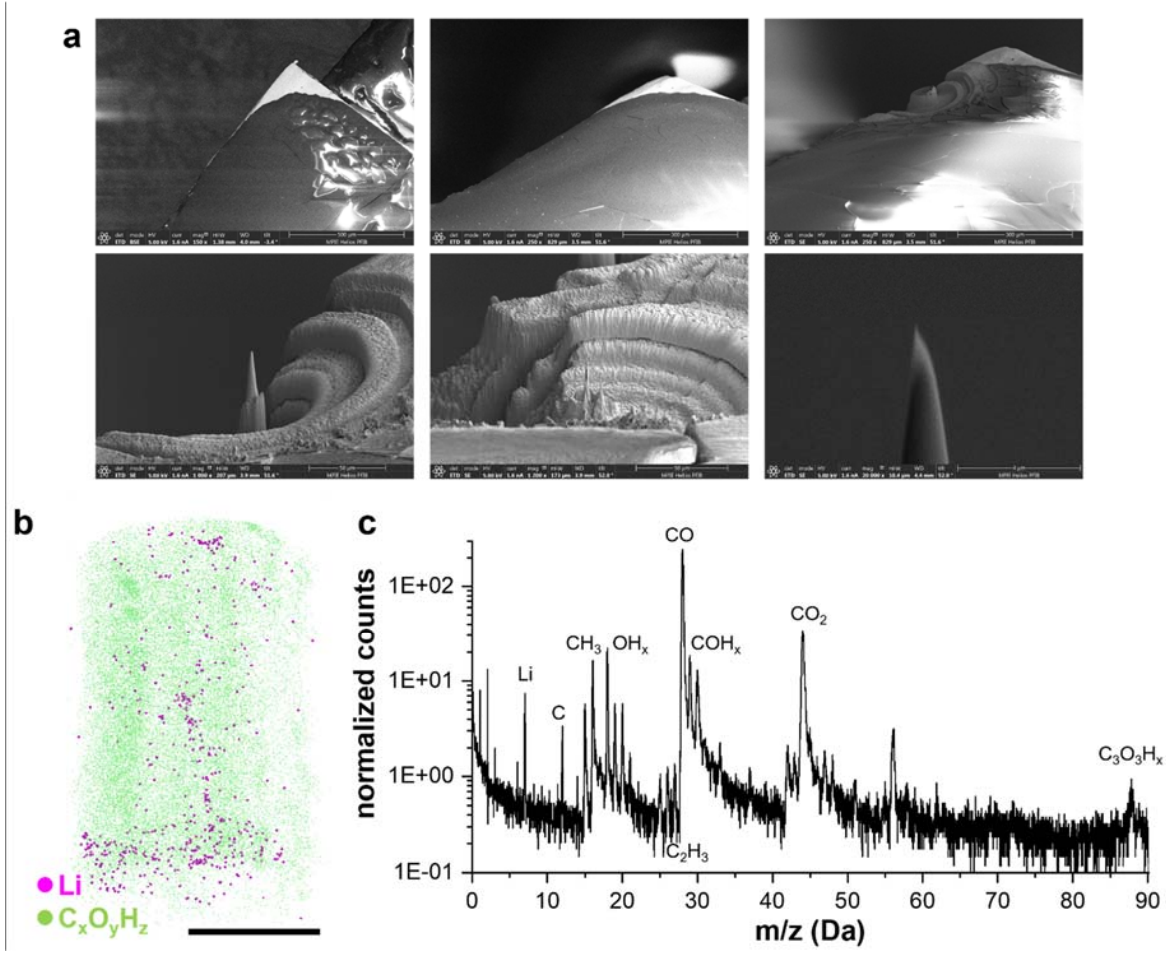

**Fig. S2.** (a) Cryo-APT specimen preparation from a raw electrolyte (non-contact with Si). (b) 3D atom map of a raw electrolyte (a scale bar = 20 nm) and (c) corresponding mass spectrum. Note that there are no Si peaks. Nano-porous gold instead of Si was used as a substrate to hold the frozen raw liquid electrolyte for cryo-APT measurement. The reconstructed Li ions (6,7 Da) are segregated locally implying that there was a phase separation of the Li salt during freezing. In the mass spectrum of the pristine (non-Si contacted) electrolyte, notable peaks at 90-85 Da are measured which originates to the $C_3O_3H_x^+$ molecular species.

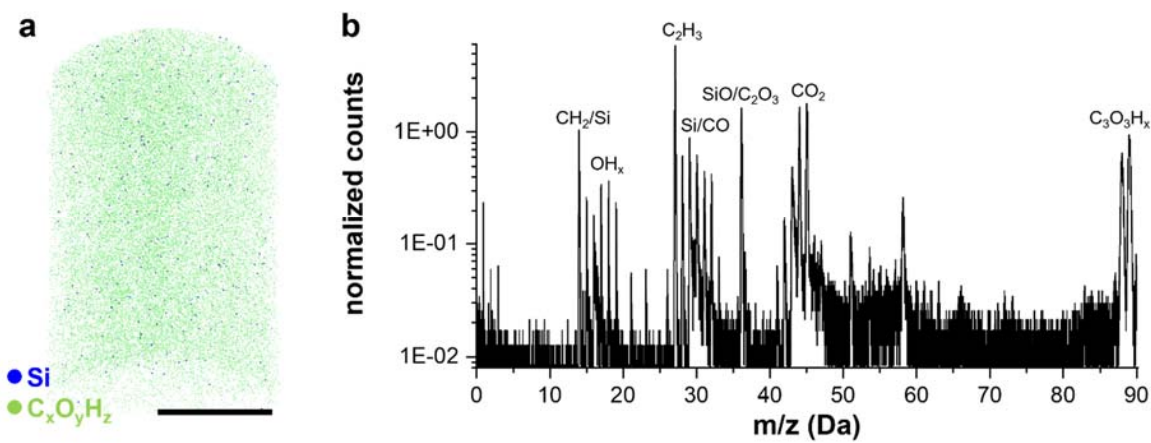

**Fig. S3.** Additional cryo-APT measurement of the 1-cycle electrolyte. (a) 3D atom map of 1-cycle electrolyte and (b) corresponding mass spectrum. Note that there is Si peaks. A scale bar is 20 nm.

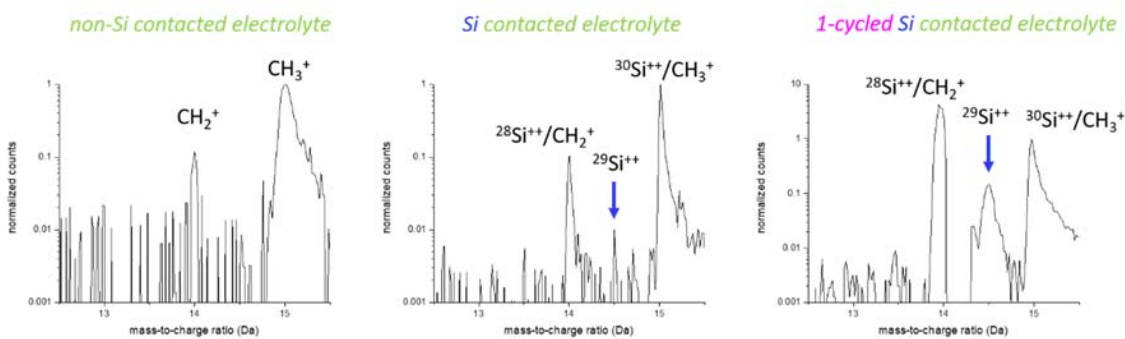

**Fig. S4.** Background corrected mass spectra of non-Si, Si-contacted and 1-cycled electrolyte. Note that peaks at 14.5 Da aren't likely from $CO^{++}$ state. The second ionization energy of CO requires 41.8 ±0.5 eV (for $CO^+$ = 14.07 ±0.05 eV)[1] whereas ionization energies of $Si^+$ and $Si^{++}$ are 8.15 and 16.34 eV. Therefore, detecting $CO^{++}$ is extremely unlikely.

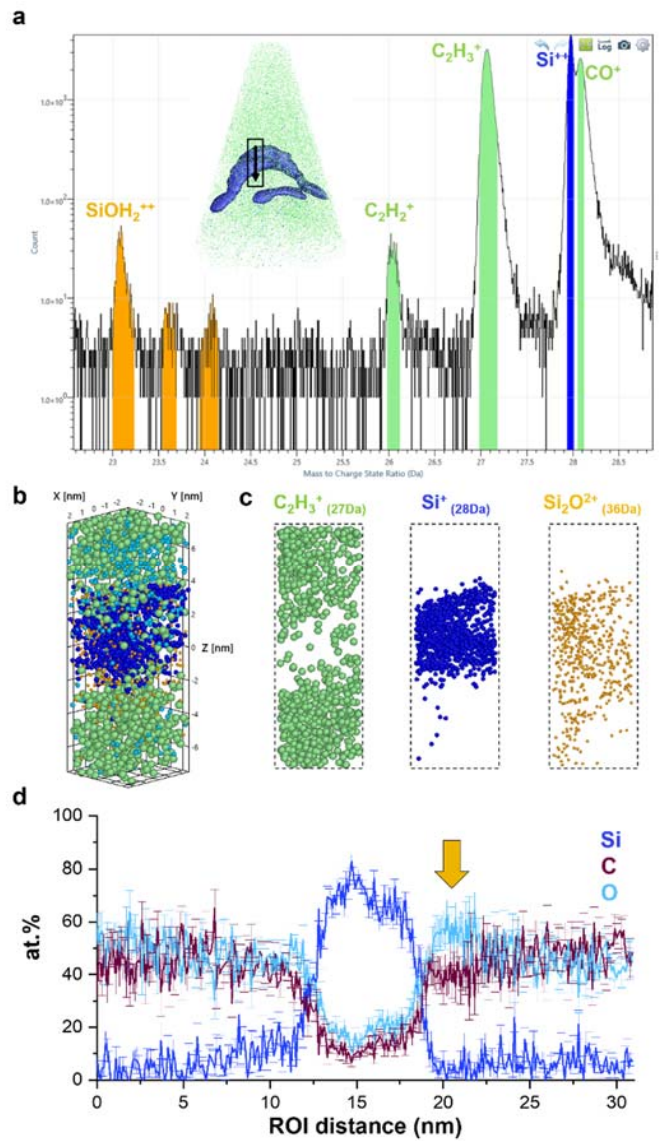

**Fig. S5.** APT analysis of 1-cycled electrolyte sample. (a) Mass spectrum of cycled electrolyte APT dataset. Note that there is a strong peak split at 28 Da originated to overlapped peaks of $CO^+$ and $Si^+$ ions, which is commonly seen in the CO measurement[2]. Inset shows that corresponding 3D atom map. (b) Extracted region of interest (5×5×15 nm³) from Fig. 2d. (c) ions distributions: $C_2H_3^+$(green), $Si^+$(blue), and $Si_2O^+$ (orange). (d) 1D compositional profiles along the Si nano-fragment in the measurement direction. Note that O enrichment at the interface can be observed (orange arrow).

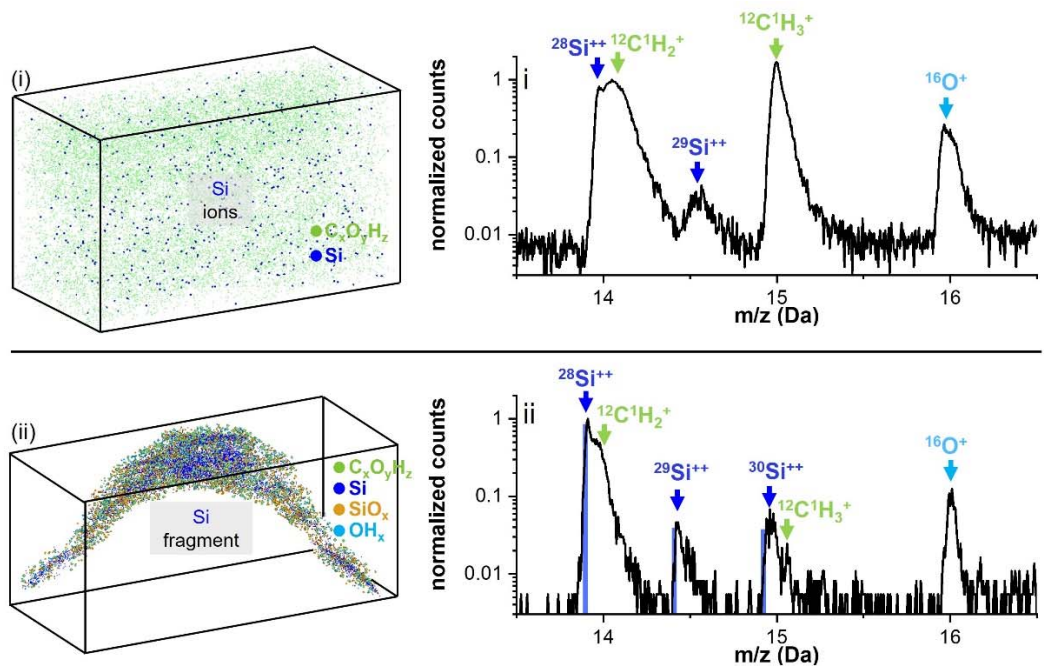

**Fig. S6.** Mass spectrum of each extracted volume from 1-cycled electrolyte APT dataset (Fig. 2d). Blue lines on the mass spectrum indicate a ratio of natural abundance isotopes of silicon.

**APT specimen and TEM lamella preparation from the 10-cycled Si anode**

After the cycling for 10 times, the Si anode was collected and rinsed with 1-methyl-2-pyrrolidone (NMP) (anhydrous 95%, Sigma Aldrich) solvent inside the $N_2$ glovebox. After removing surface residuals, it was dried in a vacuum chamber attached to the glovebox for 1 hr. Subsequently, the sample was loaded to the precision etching coating system (PECS) II (Model 685, Gatan). A 50 nm layer of Cr was deposited on the sample for surface protection. The coated sample was loaded to the dual-beam FIB (FEI Helios Nanolab 600) chamber. APT specimens from the 10-cycled sample were prepared using Ga-ion milling according to ref.[3]. Three distinguishable interest regions (topmost surface, near-surface, and bulk) were fabricated into the APT specimens. For TEM lamella, first, the surface was coated additionally with e- and Ga-ion beam-induced Pt/C layer. Then the lamella was obtained mostly following the protocol described in ref.[4].

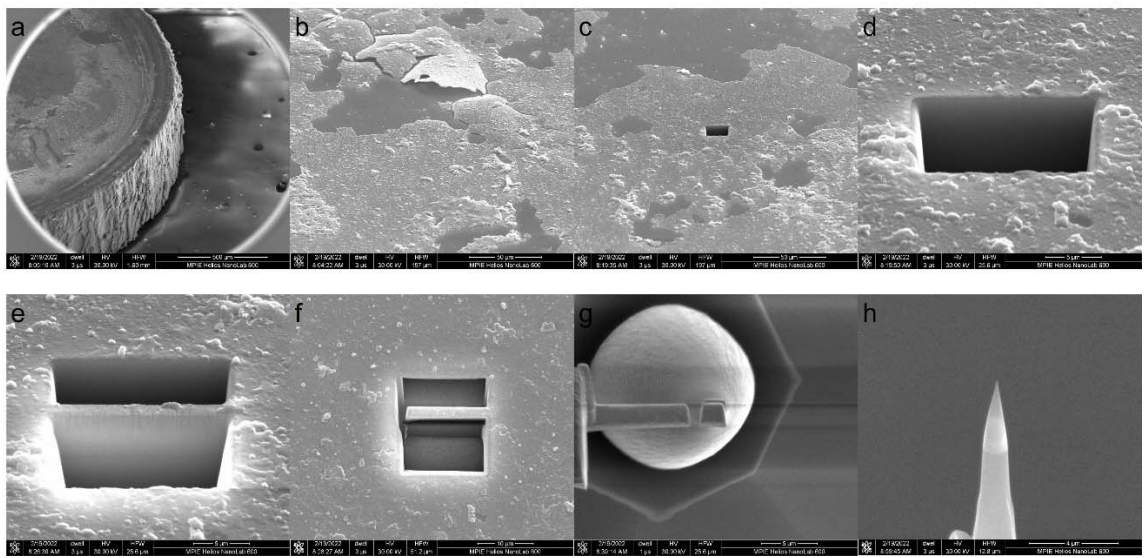

**Fig. S7.** APT specimen preparation of the 10-cycle Si anode. The disassembled electrode was coated with PECs-Cr layer (~50 nm). (a) 52°-tilted Si anode. (b) FIB/SEM surface image shows that there are delaminated layers of residuals (*i.e.* salt (please see APT/TEM results in S9 and 11)). (c)-(e) Front and back-side cuts. (f) L-shape cut to free the lamella. (g) Mounted APT sample on a commercial Si micro-post. (h) A final APT specimen.

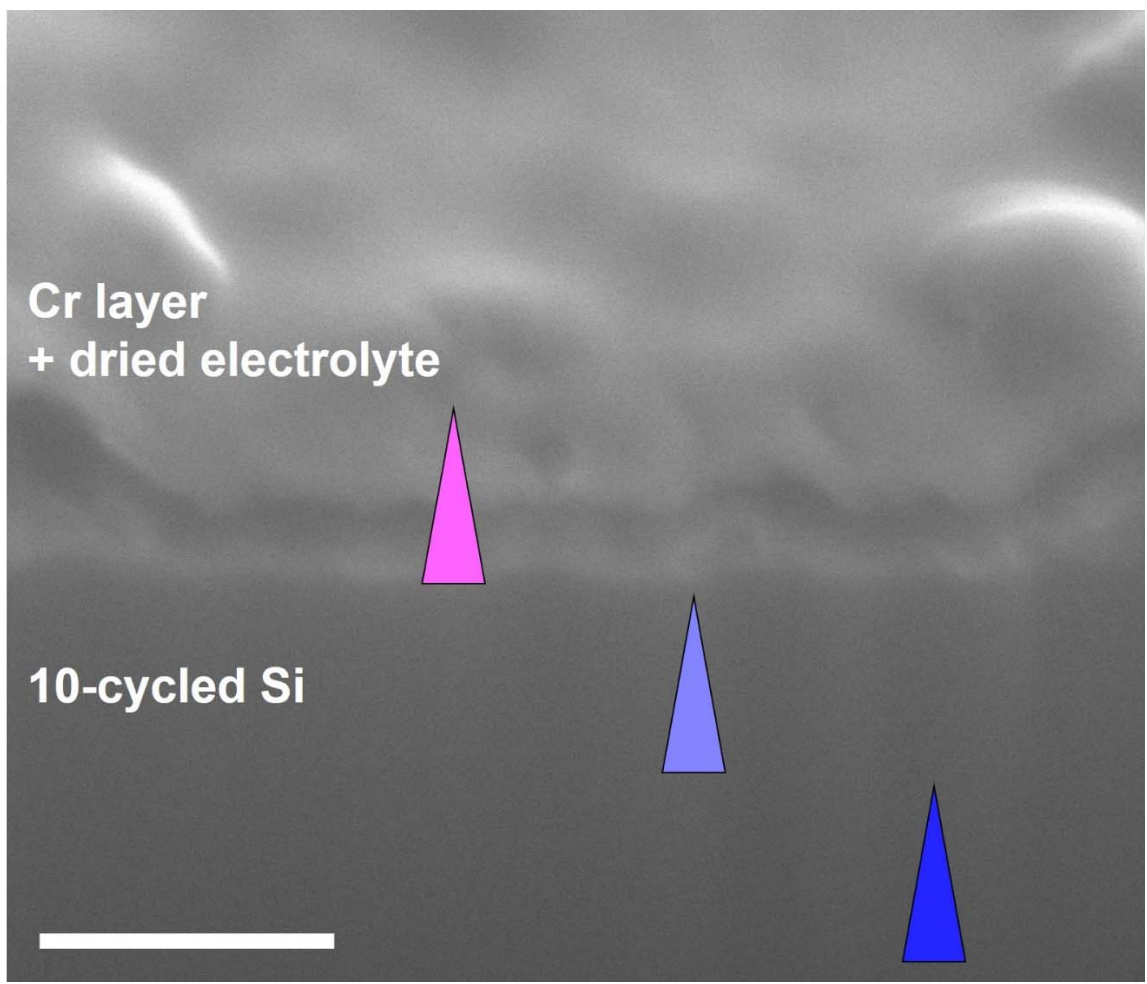

**Fig. S8.** A cross-sectional FIB/SEM image of the 10-cycled Si anode (scale bar = 500 nm). The colored triangles indicate the regions of three representative APT measurements that are presented in Fig. S9.

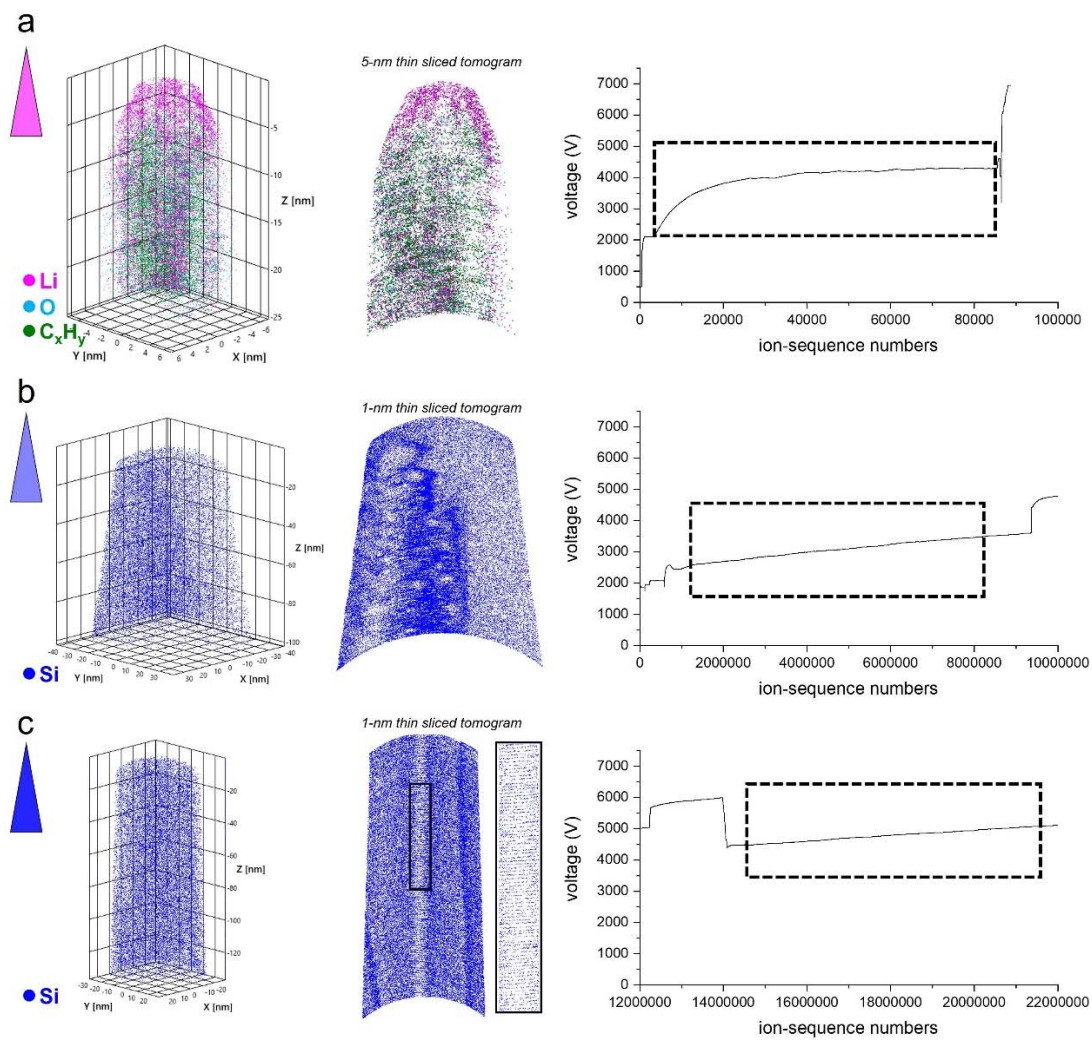

**Fig. S9.** APT analysis (3D atom map, tomogram, and V-curves) of (a) salt layer, (b) near-surface Si, (c) bottom Si from the 10-cycled Si sample. Note that ambiguous peaks in the APT dataset from the salt layer were not ranged, for instance F (LiC) *vs.* $H_3O$ or $Li_2C$ *vs.* $C_2H_2$ etc.

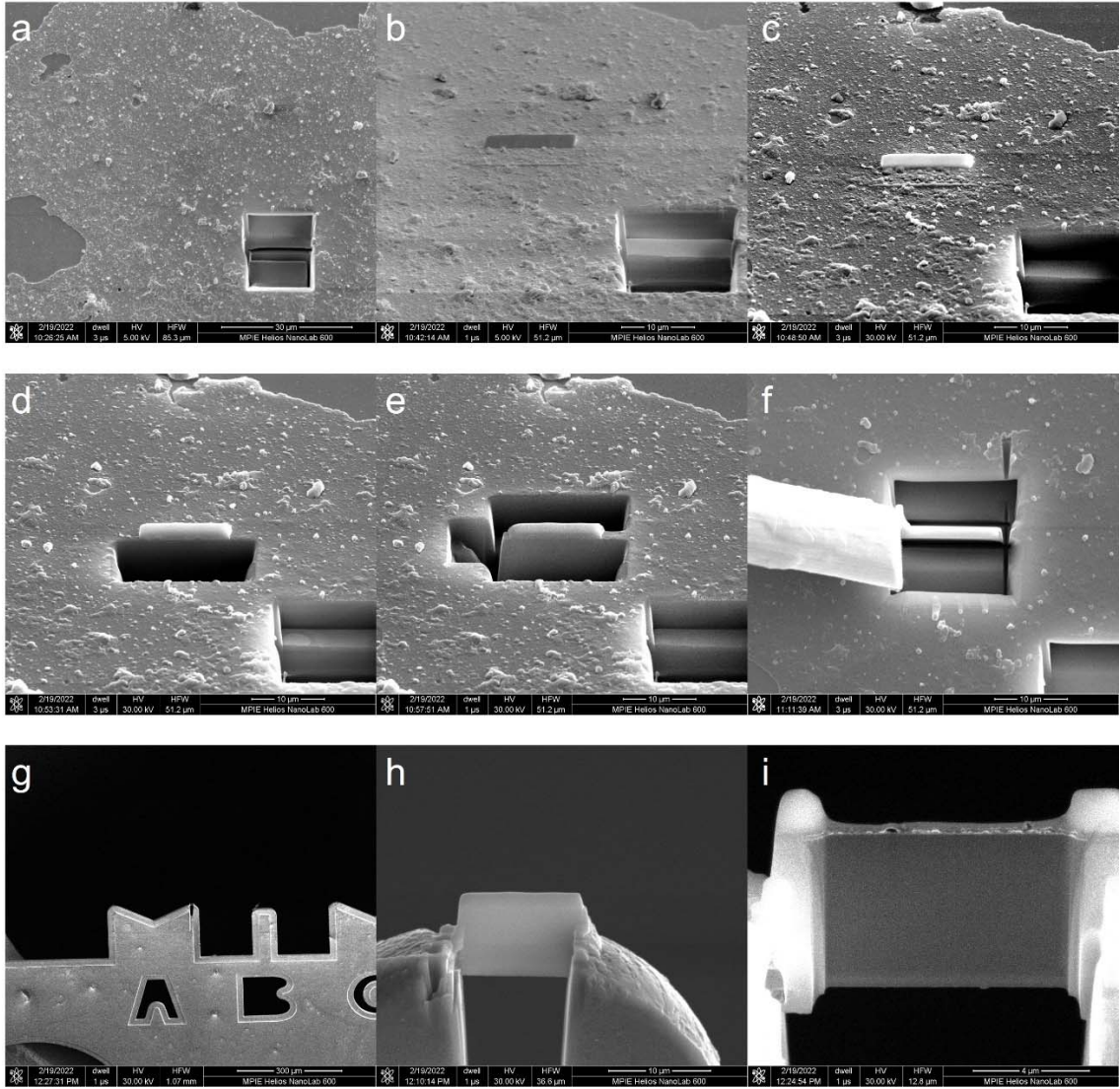

**Fig. S10.** TEM lamella preparation of the 10-cycled Si sample. (a) FIB/SEM surface image near the APT-site-lift-out region. (b) e-beam Pt/C deposition. (c) ion-beam Pt/C deposition. (d) & (e) Front-side and back-side cuts followed by the L-cut. (f) Attachment to a micro-manipulator. (g) FIB/SEM image of a commercial TEM Cu grid. (h) Mounted TEM lamella on the grid. (i) A final TEM lamella after the thinning process.

**TEM measurement**

TEM characterization of the cycled Si anode was performed in a JEM-2200FS TEM (JEOL) instrument operating at 200 kV. TEM images were acquired using a TemCam-XF416 pixelated scintillator-based complementary metal-oxide-semiconductor (CMOS) detector from Tietz Video and Image Processing Systems (TVIPS). We used Gatan Microscopy Suite® 3 Software to process TEM images, e.g. Fast-Fourier Transformation (FFT) of high-resolution TEM (HRTEM) images. The lamella specimen of the 10-cycled Si anode was titled close to Si [110] zone for the high resolution imaging of lattice fringes. Fig. S11a shows the interfacial regions between the Pt/C protection, the PECS-Cr, the electrolyte residuals, and the 10-cycled Si anode. Regions of interest have been highlighted in Fig. S11b to show local structure of the electrolyte residuals and the interface between the electrolyte residuals and the Si anodes. Fig. S12&13 present a variety of different defect structures observed in the 10-cycled Si anode.

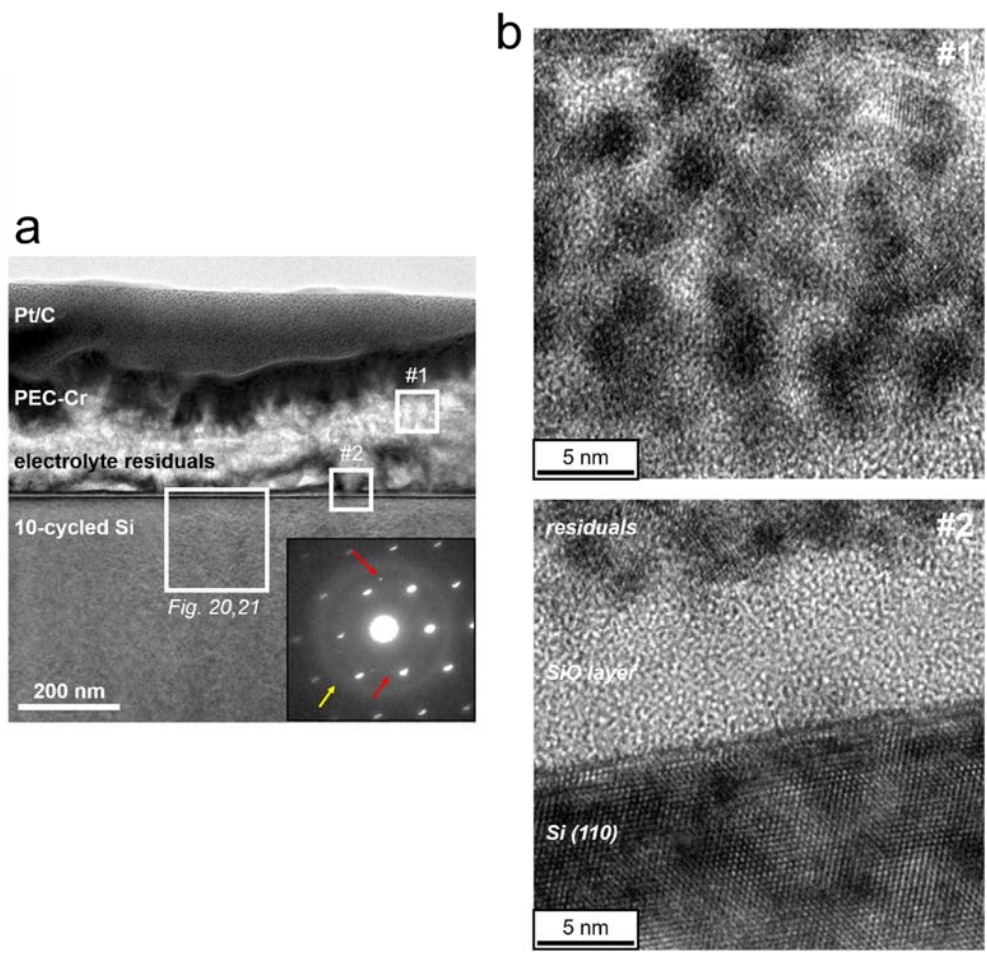

**Fig. S11.** TEM analysis of the 10-cycle Si. (a) Bright-field (BF) from the top-surface region and (b) high-resolution TEM (HRTEM) images of two small regions as marked with squares in (a). Inset image in **a** shows the complex diffraction patterns with amorphous ring (yellow) and un-identified diffraction points (red). The fast-Fourier transformation (FFT) patterns are measured along [110] Si zone axis.

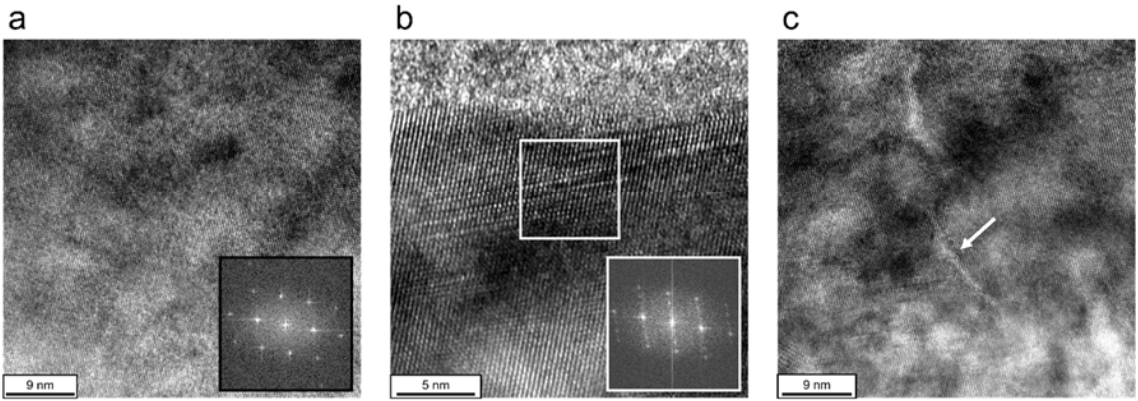

**Fig. S12.** HRTEM analysis of the 10-cycle Si. (a) Si region with a clear a single-phase Si[110] FFT pattern. (b) Defect sites at near-surface. (c) Defect sites within the Si anode.

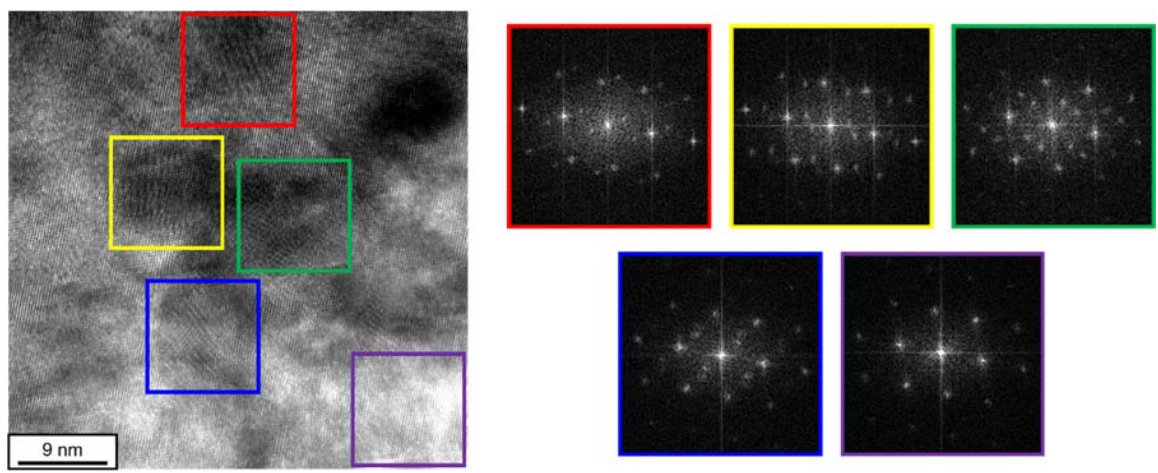

**Fig. S13.** HRTEM analysis of the 10-cycle Si. Note that these FFT patterns show different types of defects.

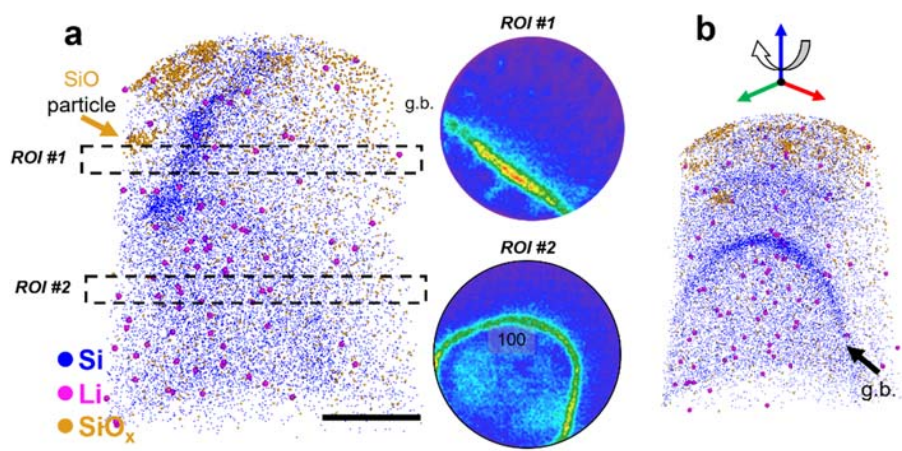

**Fig. S14.** (a) 3D atom map of the 25-cycle Si anode with Si ion density map. (b) 90°-rotated 3D atom map of 25-cycle Si anode (from Fig. 2f); a scale bar is 20 nm.

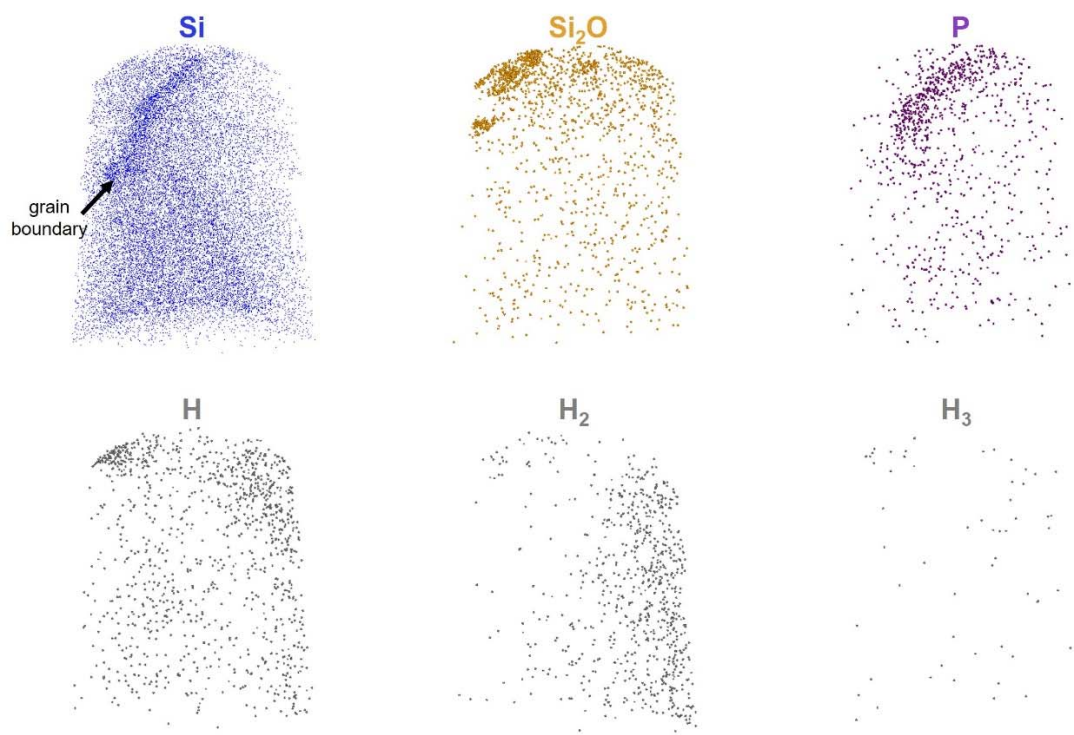

**Fig. S15.** 3D atomic distributions of Si, Si₂O, P, H, H₂, and H₃ species of the 25-cycled Si. It clearly shows that the peak at 31 Da is not related to H-Si peaks but different species (*i.e.* P). Note that O does not appear at region where P atoms locate.

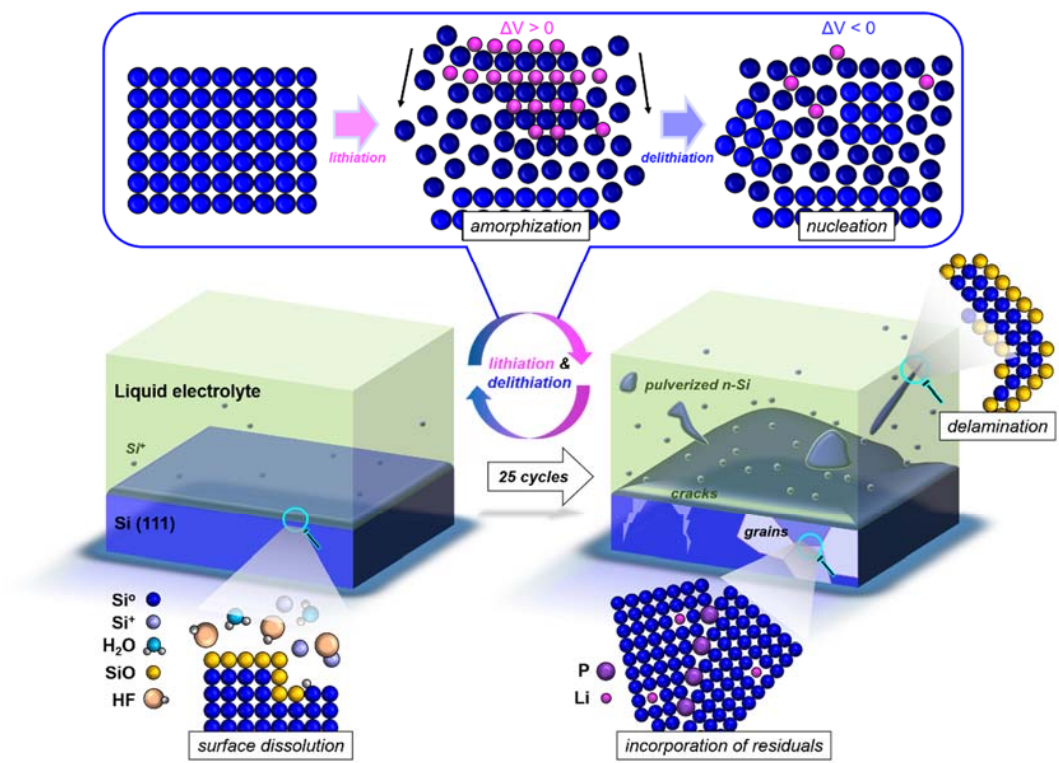

**Fig. S16.** Schematic illustration of Si anode failure: dissolution, delamination, implementation of electrolyte salt elements, and mechanical-stress relaxation.

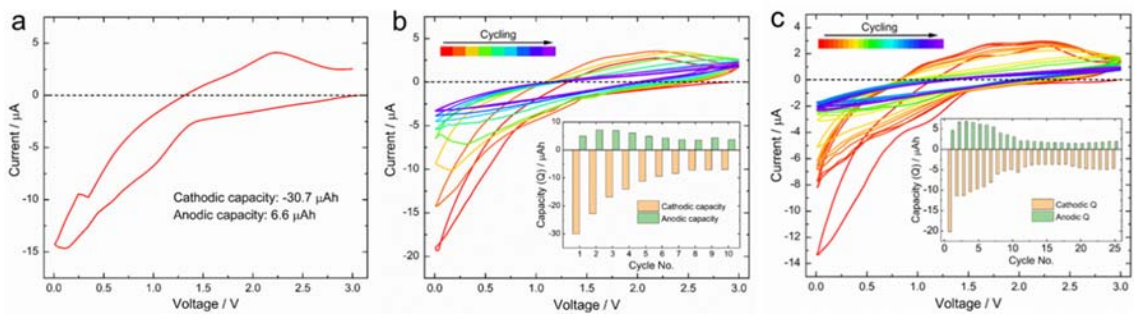

**Fig. S17.** The voltage profiles of the Li/Si cells after (a) 1, (b) 10, and (c) 25 cycles. Li/Si cells show similar electrochemical behavior during the CV scanning. From the insert figure in panel (b) and (c), evident cathodic capacity fading could be observed during the consecutive 10 and 25 cycles, respectively.

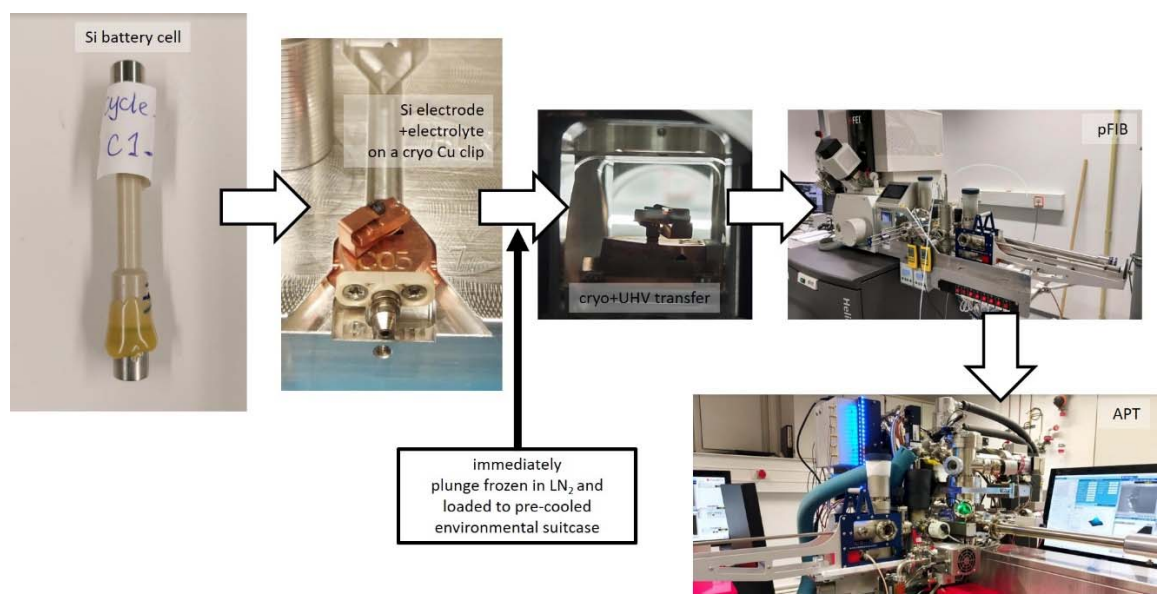

**Fig. S18.** The protocol of cryo-APT experiment. A battery cell was disassembled inside a $N_2$ glovebox and the interested cell was mounted on the Cu clip and immediately quenched into $LN_2$. Subsequently, the clip was loaded to the pre-cooled UHV carry suitcase (-190 °C and $10^{-9}$ mbar) and transferred to Gatan cryo-stage installed plasma-FIB. After final milling, the cold specimen was transferred back to the suitcase maintaining cryo-UHV conditions and was detached from the PFIB and mounted onto a LEAP 5000 XS atom probe system. Finally, the puck was transferred under cryo-UHV conditions to the atom probe analysis chamber. Details on the specific home-made installation are described in Ref.[5].

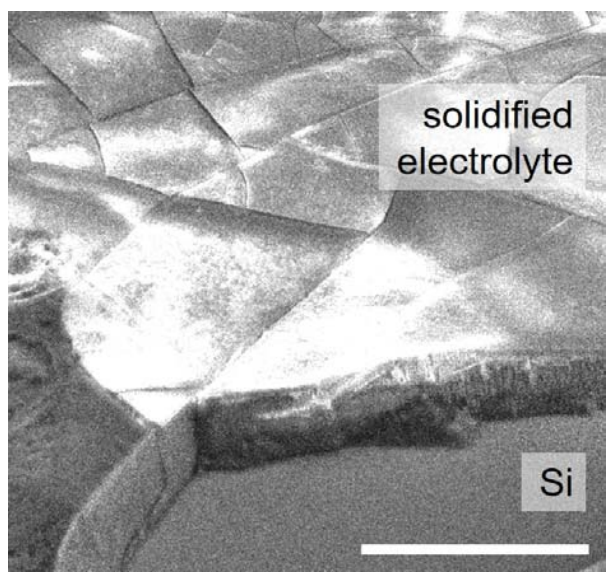

**Fig. S19.** Xe-plasma FIB/SEM image of frozen liquid electrolyte on the 1-cycle Si electrode. A white scale bar is 100 μm.

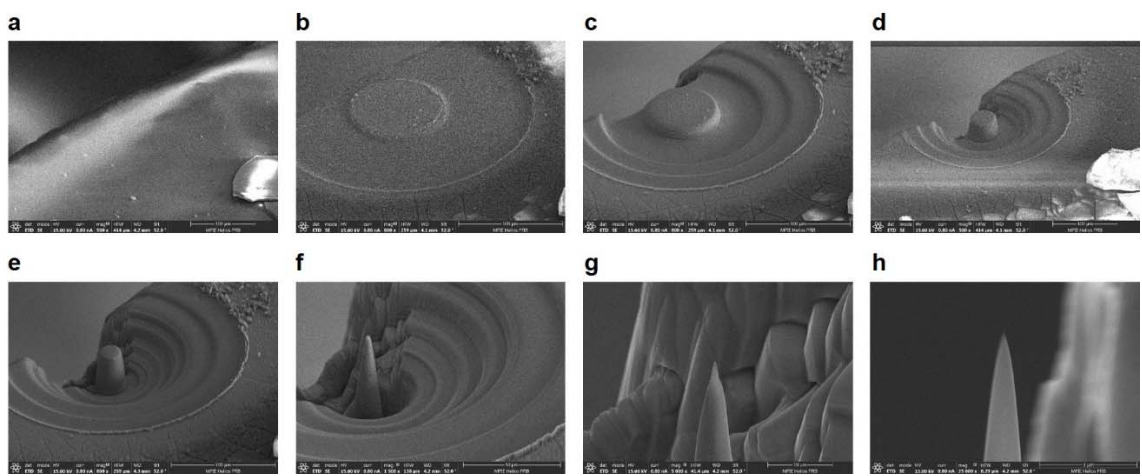

**Fig. S20.** APT specimen preparation of frozen liquid electrolyte/cryo-Si sample. The Halpin protocol was adapted to obtain a pillar shape[6]. (a) First, an ion-beam-circle pattern of outer diameter of 200 μm and inner diameter of 100 μm was set at 30 kV and 1.3 μA for 10 min. (b)-(e) Then patterns of outer and inner diameters were gradually reduced until inner diameters reach 30 μm with a depth of 50 μm. (f) The ion-beam current was set at 60 nA and the pillar was milled with a circle pattern (outer/inner = 50/10 μm) further down to fabricate into a toblerone shape. (g) Once a typical APT specimen geometry was obtained, the ion-beam current was reduced to 1 nA. (f) The final milling process was done at ion-beam current of 0.3 nA.

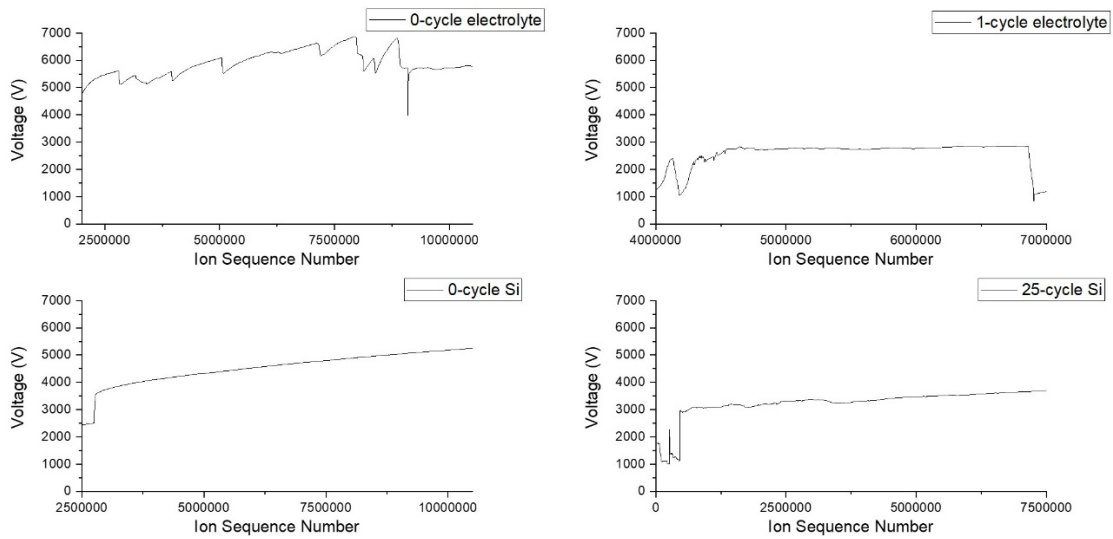

**Fig. S21.** Voltage curve of each cryo-measurement: 0-cycle electrolyte, 0-cycle Si anode, 1-cycle electrolyte, 1-cycle electrolyte containing n-Si debris, 25-cycle Si anode.